\documentclass[11pt]{article}
\usepackage{amsmath,amssymb,amsthm}
\usepackage{tikz}
\usepackage{graphicx}
\usepackage{changes}
\newtheorem{proposition}{Proposition}
\theoremstyle{definition}

\begin{document}

\title{Hierarchical Incentives and the Evolution of Local Cooperation in Wartime: A Continuous Strategy Approach}
\author{Leonardo Becchetti\\University of Tor Vergata in Rome\and
Francesco Salustri\\
Roma Tre University \& University College London
\and
Nazaria Solferino\\
Department of Economics, Statistics and Business\\
Universitas Mercatorum, Rome
}

\date{}
\maketitle

\begin{abstract}
Historical episodes such as the World War I ``live-and-let-live'' system and the Christmas Truce of 1914 demonstrate that opposing military units can establish spontaneous, local cooperation even in extreme conflict environments. Such cooperative behavior is typically fragile and temporary, while large-scale wars persist. We develop a hierarchical decision problem in which local units adopt contingent strategies that depend on interactions, accumulated payoffs, and signals from a central command. The command authority can impose enforcement that penalizes non-aggression to prolong hostilities. Our model features a continuous space of parametric strategies and formalizes replicator dynamics over the population. We analytically characterize the conditions under which local cooperation emerges as a stable evolutionary equilibrium and identify critical thresholds of central enforcement that destroy cooperative equilibria. We show that stable peace requires either alignment of command incentives with frontline welfare, external constraints on enforcement, or diminishing political returns to conflict. The framework provides a micro-founded explanation for the persistence of war despite locally beneficial cooperation.

\textbf{JEL Codes:} C73, D74, D78, H56, Z13

\textbf{Keywords:} Evolutionary games, Repeated interactions, Military conflict, Hierarchical incentives, Cooperation, Truce dynamics
\end{abstract}

\section{Introduction}
The emergence of cooperative behavior among adversaries in situations of extreme conflict poses a fundamental question in economics, political science, and evolutionary game theory: why do local agents sometimes choose to cooperate, even when large-scale incentives favor continued aggression? Historical accounts of World War I provide compelling examples. Soldiers on the Western Front frequently developed informal arrangements to reduce hostilities, a phenomenon described by \cite{Ashworth} as the ``live-and-let-live'' system. These arrangements included tacit truces, the avoidance of unnecessary artillery fire, and limited coordination to recover the wounded. Perhaps the most famous example is the Christmas Truce of 1914, during which German and Anglo-French units met in ``no-man’s-land'' to exchange gifts, sing carols and temporarily suspend fighting. These events illustrate that cooperation can emerge spontaneously, even under conditions of extreme risk and severe hierarchical pressure to maintain combat operations.  
Such observations create a theoretical puzzle. Standard models of war, conflict, and strategic interactions often predict persistent aggression whenever individual or state-level incentives favor fighting (\cite{Fearon};\cite{Powell}). Similarly, classical repeated game theory (\cite{Fudenberg};\cite{ Axelrod}) demonstrates that cooperation can be sustained in iterated interactions through reciprocity. These models typically assume that agents share the same objective function and do not account for hierarchical manipulation of payoffs. In wartime, this assumption is violated: higher-level authorities often derive independent benefits from continued conflict, whether political, organizational, or strategic, creating a fundamental tension between local cooperation and centralized objectives.

\cite{Axelrod} seminal work demonstrated that simple reciprocal strategies, such as Tit-for-Tat, can generate stable cooperation in the Iterated Prisoner's Dilemma. Subsequent studies (\cite{Nowak}; \cite{Hofbauer}\added{)} extended these insights to stochastic strategies, continuous strategy spaces, and evolutionary dynamics, showing how cooperative behavior can emerge through selection in populations. However, these models largely abstract away from hierarchical structures, assuming that all agents face the same incentives and that payoffs are exogenous and fixed. The presence of a command authority that can manipulate payoffs or impose penalties for non-aggressive behavior significantly complicates the strategic environment and can undermine the sustainability of local cooperation.

The central research question of this paper is:\\ \textit{under what conditions can local cooperative behavior in wartime become a stable evolutionary equilibrium, and how do hierarchical incentives from command authorities influence the emergence, stability, or collapse of such cooperation?}\\ 
Addressing this question is critical for understanding the persistence of war despite the presence of mutually beneficial cooperation at the micro level. The question is not only theoretically interesting but also of practical relevance: insights into the conditions that sustain local peace can inform military strategy, rules of engagement, and policy interventions aimed at conflict mitigation.

This paper introduces a novel hierarchical evolutionary framework to answer this question. Frontline units are modeled as agents adopting parametric, stochastic strategies taking into account command signals into decisions of truce or attack. The strategy space is continuous, allowing for nuanced behaviors beyond simple binary choices.\\ A central authority represents hierarchical command, which can impose \emph{enforcement} that penalizes non-aggression, reflecting realistic incentives for commanders to prolong conflict. The model incorporates \emph{replicator dynamics} over continuous strategy spaces, capturing how strategies evolve over time as agents adapt to relative fitness outcomes. This structure allows for the formal analysis of equilibrium distributions of strategies under varying conditions of enforcement, reciprocity, risk aversion, and command pressure.

The paper makes three key contributions. First, it formalizes the interplay between local cooperative dynamics and centralized enforcement in a hierarchical, evolutionary game-theoretic model, bridging a gap between the descriptive historical literature and formal economic modeling. Second, it identifies analytically critical thresholds of enforcement beyond which local cooperation collapses, providing clear, testable conditions for when micro-level peace can or cannot persist. Third, it connects these theoretical insights to practical considerations, showing that stable local cooperation requires either: (i) alignment of hierarchical incentives with the welfare of frontline units, (ii) external institutional or legal constraints on the ability of commanders to punish truce behavior, or (iii) diminishing political returns to prolonged conflict. These findings offer a micro-founded explanation for why wars can persist despite the existence of opportunities for mutually beneficial cooperation at the front line.

This research is innovative in several ways. Unlike classical repeated or evolutionary game models, it explicitly incorporates \emph{hierarchical incentives} that can distort local payoffs. Unlike standard models of war in political economy, which typically focus on state-level bargaining or commitment problems (\cite{Fearon},\cite{Powell}), it analyzes micro-level agent interactions embedded within a hierarchical structure. Finally, it directly links theoretical results to historical observations, providing a formal framework that can explain why phenomena such as the Christmas Truce and ``live-and-let-live'' arrangements are both possible and inherently fragile.

In addition to its theoretical contribution, the model has important implications in the real-world. By formalizing the conditions under which local cooperation can survive hierarchical pressure, it offers guidance in designing rules of engagement, institutional monitoring mechanisms, and command incentives that could mitigate unnecessary violence in contemporary conflicts. It also provides a foundation for analyzing related phenomena such as dissident coordination in authoritarian regimes or cooperative behavior in competitive organizational environments, where central authority may actively resist locally beneficial cooperation.

\section{Literature Review}

The study of cooperation and conflict has been approached from multiple disciplines, including evolutionary biology, economics, political science, and theoretical game theory. A foundational contribution in this area is the work of \cite{Axelrod}, who demonstrated how cooperation can arise among self-interested agents through repeated interaction in the Iterated Prisoner’s Dilemma \cite{Axelrod}. Axelrod's findings have shaped a vast literature on reciprocal strategies, social norms, and stability of cooperative behaviour.

Evolutionary game theory (EGT) extends these insights by considering populations of interacting agents adapting over time through imitation or selection processes \cite{Sandholm2010, Traulsen2023}. Early work by Nowak and Sigmund highlighted mechanisms such as indirect reciprocity and spatial structure that enhance cooperation \cite{Nowak, Nowak1998, Nowak2004}. Within this framework, biological concepts like reciprocal altruism and evolutionary stable strategies (ESS) were formalized, supported by John Maynard Smith’s theory of evolution and games \cite{MaynardSmith1973}.

EGT has also been applied to structured populations and multi-group interactions. Studies on network reciprocity demonstrate that cooperation can persist in structured environments where agents interact locally rather than randomly \cite{Ohtsuki2006}. Reward and punishment mechanisms have been shown to significantly influence cooperative dynamics, revealing the complex role of incentives in stabilizing cooperation \cite{Fehr2000}, \cite{Fudenberg}. Recent work emphasizes the importance of community structure and learning mechanisms for robust cooperation \cite{Powell}, \cite{Fudenberg}.

In parallel, cooperation has been studied in contexts of competition and conflict. Evolutionary analyses of competitive games like Hawk–Dove highlight conditions under which aggression or peaceful coexistence may prevail \cite{Hofbauer}. Research exploring mixed cooperative–competitive environments finds that inter-group competition can affect within‑group cooperation levels \cite{Javarone}. These results provide partial insight into how conflict influences cooperative behaviour, a crucial consideration for models of war and contest dynamics.

Applying EGT to real‑world resource conflicts, such as water management disputes, has yielded models showing how institutional interventions can shift outcomes toward cooperation or non‑cooperation \cite{Cao}. These tripartite evolutionary game frameworks illustrate that external agents—institutional enforcers—can play a decisive role in sustaining or undermining cooperation, a theme directly relevant to hierarchical incentives in wartime.

A growing body of recent literature focuses on continuous strategy spaces and information processing capabilities in cooperation. For example, some works investigate how diversity in strategy tendencies impacts evolutionary outcomes. Emerging papers also explore comprehensive strategy spaces beyond hand‑picked sets, revealing hidden cooperation pathways \cite{Song}.

There is also increasing interest in information asymmetry and adaptive analysis capabilities. Differential information processing has been shown to promote altruistic behaviour in evolutionary settings \cite{Wang}. These modern developments expand classical models by allowing richer behavioural states and nuanced adaptation.

Despite this breadth, most existing literature either focuses on abstract cooperation in stylized games or on conflict resolution at the macro (state‑to‑state) level \cite{Fearon};\cite{Powell}, leaving gaps in how microscale interactions under hierarchical enforcement incentives shape cooperation. Few formal models integrate central authority interventions, asymmetric payoffs, and continuous strategy evolution in conflict environments, making it difficult to systematically understand stable truce formation in wartime.

Several related studies illustrate components of the problem. Economic models of enforcement indicate that coercive punishment can sustain cooperation under some structures \cite{Acemoglu}, while institutional incentives can be deployed to maximize coordination participation in complex dilemmas \cite{Ogho}. Recent work on punishment in bipartite societies shows how inter‑group sanctioning affects cooperation outcomes, highlighting the importance of structural roles in evolutionary dynamics \cite{Feng}.

In summary, while evolutionary game theory offers powerful tools for understanding cooperative behaviour and conflict, and while there is an expanding literature on structured populations, incentives, and mixed cooperation–competition environments, there remains a theoretical gap regarding models that combine hierarchical enforcement incentives with local evolutionary dynamics in conflict contexts. The present work aims to fill this gap by incorporating continuous strategy evolution, centralized enforcement, and replicator dynamics to determine conditions under which local cooperation emerges and persists in the presence of hierarchical incentives.

On our knowledge, despite the extensive body of work on evolutionary game theory, cooperation, and conflict, a significant gap remains in understanding how hierarchical incentives and top-down authority affect the emergence and stability of cooperation in conflict situations. Most classical models, including Axelrod's seminal contributions \cite{Axelrod, AxelrodHamilton1981}, focus on decentralized interactions, repeated games, or symmetric populations. Even recent studies on structured populations, continuous strategy spaces, and dynamic reward or punishment mechanisms \cite{Zhou, Lyu, Geng} largely neglect the role of agents with explicit command power whose preferences may favor prolonging conflict.

Empirical and institutional studies \cite{Cao, Feng, Fearon, Powell} explore cooperation and conflict resolution at the macro level but fail to model microscale evolutionary dynamics under hierarchical control, where local agents might cooperate temporarily while overarching authority manipulates incentives to sustain conflict. This creates a tension between local cooperation and global conflict persistence, which existing models do not capture.

Furthermore, the majority of current literature either assumes homogeneous or exogenously fixed payoffs or studies cooperation in abstract settings without directly linking to real-world conflict environments. There is therefore no formal evolutionary model that integrates continuous strategy evolution, local reciprocity, and top-down hierarchical incentives in conflict scenarios, preventing a systematic analysis of the conditions under which stable truce or peace equilibria could emerge.

The present work addresses this gap by developing a novel continuous-strategy evolutionary game model that explicitly incorporates local strategy adaptation through replicator dynamics, hierarchical enforcement incentives, and asymmetric payoff structures reflecting command preferences. This framework allows for an analysis of when and how local cooperation can persist or collapse in the presence of central authority incentives favoring conflict, providing both theoretical insight and policy-relevant guidance for stabilizing cooperation in real-world conflict situations.

\section{The Model}

We consider a population of frontline units engaged in repeated bilateral interactions during a conflict. Time is continuous and indexed by $t \geq 0$. Interactions are decentralized and occur between randomly matched agents from two opposing populations, which we assume to be symmetric. In each encounter, agents choose whether to engage in combat or to refrain from fighting, the latter interpreted as a local truce.

\subsection{Strategies and Behavior}

Each agent is characterized by a continuous strategy parameter $\theta \in \mathbb{R}$, which determines the agent's propensity to cooperate (i.e., refrain from fighting). Let $p(\theta)$ denote the probability that an agent with type $\theta$ chooses Truce ($T$) in a given interaction. We assume a logistic probabilistic choice function:
\begin{equation}
p(\theta) = \frac{1}{1 + e^{-\theta}} \in (0,1),
\end{equation}
which is smooth, strictly increasing, and satisfies $\lim_{\theta \to -\infty} p(\theta) = 0$ and $\lim_{\theta \to +\infty} p(\theta) = 1$.

The state of the population at time $t$ is described by a probability density function $f(\theta,t)$ over $\Theta = \mathbb{R}$, with $\int_{-\infty}^{+\infty} f(\theta,t)\, d\theta = 1$ for all $t$.

\subsection{Payoffs and the Nature of the Game}

When two agents meet, each independently chooses either Truce ($T$) or Attack ($A$). Let $R$ be the payoff when both choose Truce, $V$ the payoff for an attacker meeting a truce‑maker, $S$ the sucker's payoff for a truce‑maker meeting an attacker, and $P$ the payoff when both attack. The material payoff matrix is
\[
\begin{array}{c|cc}
 & T & A \\
\hline
T & R, R & S, V \\
A & V, S & P, P
\end{array}
\]

In the classic Prisoner's Dilemma we would have $V > R > P > S$, which implies that mutual defection is the unique Nash equilibrium. However, such a setting would leave little room for the emergence of local truces without exogenous enforcement. To capture the tension between hierarchical incentives and decentralized cooperation, we instead consider a coordination game where both pure strategies can be stable under appropriate conditions. Specifically, we assume
\begin{equation}
R > V \quad \text{and} \quad P > S,
\end{equation}
so that mutual cooperation (Truce, Truce) and mutual defection (Attack, Attack) are both Nash equilibria. More specifically, under
\eqref{eq:coord}, each agent prefers $T$ if the opponent plays $T$ (since
$R>V$) and prefers $A$ if the opponent plays $A$ (since $P>S$). This makes the game a stag-hunt or assurance game, which differs from the Prisoner's Dilemma as cooperation is individually rational, making the enforcement threshold $R-V$ the key parameter that determines whether peace survives hierarchical pressure. This assumption reflects situations in which soldiers may coordinate on either fighting or refraining, depending on expectations and external pressures. In other words, it reflects the idea that, after sustained contact, frontline soldiers derive a higher material and psychological utility from mutual restraint than from a unilateral attack on a passive opponent, and that even mutual combat is preferable to being
the only party to suffer losses. It also ensures that the thresholds that will appear in the analysis are well-defined and meaningful. 

In addition to the material payoffs, agents are subject to hierarchical enforcement. Let $e \geq 0$ denote the intensity of enforcement. If an agent chooses truce, a penalty proportional to $e$ is imposed.

Denote by $\bar{p}(t) = \int_{\mathbb{R}} p(\theta) f(\theta,t) d\theta$ the average probability of truce in the population. The expected payoff of an agent with strategy $\theta$ is
\begin{align}
\Pi(\theta,\bar{p},e) &= p(\theta)\big[ \bar{p} R + (1-\bar{p}) S \big]
+ (1-p(\theta))\big[ \bar{p} V + (1-\bar{p}) P \big] - e\,p(\theta) \nonumber\\
&= \underbrace{\bar{p} V + (1-\bar{p}) P}_{\pi_A(\bar{p})}
+ p(\theta) \underbrace{\big[ \bar{p}(R-V) + (1-\bar{p})(S-P) - e \big]}_{B(\bar{p},e)}.
\end{align}
Thus $\pi_A$ is the payoff of an agent who always attacks, and $B(\bar{p},e)$ is the net benefit of choosing truce over always attacking.

\subsection{Evolutionary Dynamics}

Strategies evolve according to the continuous‑time replicator equation over the strategy space:
\begin{equation}
\frac{\partial f(\theta,t)}{\partial t} = f(\theta,t)\big[ \Pi(\theta,\bar{p}(t),e) - \bar{\Pi}(t) \big],
\end{equation}
where $\bar{\Pi}(t) = \int_{\mathbb{R}} \Pi(\theta,\bar{p}(t),e) f(\theta,t) d\theta$ is the average payoff.

\subsection{Aggregate Dynamics}

Because payoffs depend on $\theta$ only through $p(\theta)$, the dynamics of the mean cooperation level $\bar{p}$ can be derived. Using the fact that under the replicator dynamics the variance of $p$ satisfies $Var_\theta[p] = \bar{p}(1-\bar{p})$ for a broad class of distributions, we obtain
\begin{equation}
\dot{\bar{p}} = \bar{p}(1-\bar{p})\, B(\bar{p},e).
\end{equation}
Substituting $B(\bar{p},e)$ and collecting terms gives
\begin{equation}
\dot{\bar{p}} = \bar{p}(1-\bar{p})\Big[ \Delta \bar{p} + (S-P) - e \Big],
\label{eq:main}
\end{equation}
where we define
\[
\Delta = (R-V) - (S-P).
\]
Under our assumptions $R-V > 0$ and $P > S$, we have $S-P < 0$, hence $\Delta > 0$. This implies that any interior equilibrium (if it exists) is unstable, while the pure equilibria may be stable depending on the enforcement level.

\subsection{Equilibria and Stability}

The stationary points of \eqref{eq:main} are obtained by setting $\dot{\bar{p}}=0$:
\[
\bar{p}^*_0 = 0,\qquad \bar{p}^*_1 = 1,\qquad \bar{p}^*_{\!int} = \frac{e + P - S}{\Delta}.
\]
Because $P-S > 0$, the numerator is positive for all $e \ge 0$. The interior equilibrium lies in $(0,1)$ exactly when
\begin{equation}
0 < \frac{e + P - S}{\Delta} < 1 \quad \Longleftrightarrow \quad e < \Delta - P + S = R-V.
\end{equation}
Thus for $e < R-V$, there exists an interior equilibrium; for $e \ge R-V$, it either coincides with $\bar{p}=1$ or lies above it, and we effectively have only the two boundary equilibria.

To determine stability, we linearize. Let $g(\bar{p}) = \bar{p}(1-\bar{p})[\Delta \bar{p} + (S-P) - e]$. Then
\[
g'(\bar{p}) = (1-2\bar{p})[\Delta \bar{p} + (S-P) - e] + \bar{p}(1-\bar{p})\Delta.
\]

\begin{itemize}
\item At $\bar{p}=0$: $g'(0) = (S-P) - e$. Since $S-P < 0$ and $e \ge 0$, this expression is always negative. Hence $\bar{p}=0$ is locally stable for all $e \ge 0$.
\item At $\bar{p}=1$: $g'(1) = -[\Delta + (S-P) - e] = -(R-V - e)$. Therefore,
  \[
  g'(1) < 0 \iff e < R-V, \quad g'(1) > 0 \iff e > R-V.
  \]
  Hence $\bar{p}=1$ is stable only when $e < R-V$; it becomes unstable when $e$ exceeds this threshold.
\item At $\bar{p}^*_{\!int}$ (when it exists, i.e. $0 < e < R-V$), the first term vanishes because the bracket is zero, leaving
  \[
  g'(\bar{p}^*_{\!int}) = \bar{p}^*_{\!int}(1-\bar{p}^*_{\!int})\Delta > 0,
  \]
  since $\Delta > 0$. Thus the interior equilibrium is unstable whenever it exists.
\end{itemize}

In summary, under the coordination game assumptions ($R > V$, $P > S$) we have the following regimes:
\begin{itemize}
\item For $0 \le e < R-V$: two stable pure equilibria ($\bar{p}=0$ and $\bar{p}=1$) coexist, separated by an unstable interior equilibrium. The population may converge to either depending on initial conditions.
\item For $e > R-V$: only $\bar{p}=0$ remains stable; $\bar{p}=1$ loses stability and the interior equilibrium no longer lies in $(0,1)$. The system is monostable with full conflict as the unique attractor.
\end{itemize}

The set of all starting points of initial cooperation from which the system will eventually converge to $\bar{p}=1$ (i.e., full peace) shrinks as $e$ increases toward $R-V$, since $\bar{p}^*_{\!int} \to 1$ as $e \to R-V$. In other words, if the population begins with a proportion of cooperators above a certain threshold, the evolutionary dynamics will carry it all the way to universal truce. If it starts below that threshold, it slides toward full conflict instead. Commanders who
want to tip the population from peace to conflict need not set the intensity of enforcement $e$ far
above the threshold $R-V$, which becomes especially policy-relevant, as small increases in
enforcement near the threshold can have discontinuous effects on long-run cooperation. This captures the intuitive notion that hierarchical pressure can undermine local cooperation even in a coordination game where peace would otherwise be sustainable.

\subsection{Role of the Assumptions and Possible Extensions}

The choice of a coordination game is motivated by the desire to model situations where both peace and conflict are self‑reinforcing social conventions, and where external enforcement can tip the balance. This contrasts with the Prisoner's Dilemma, where conflict is the only equilibrium. Therefore, we maintain the coordination game specification as it provides a clear and intuitive benchmark for studying the strategic use of enforcement by a command authority.
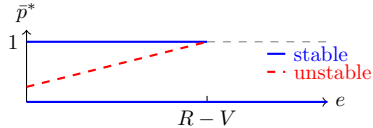
\begin{figure}[htbp]
\centering
\begin{tikzpicture}[scale=0.8, transform shape, every node/.style={font=\small}]
    % Parametri (valori illustrativi)
    \def\RminusV{3}
    \def\SminusP{-1}  % S-P è negativo
    \def\ymax{1.2}
    \def\xmax{5}
    
    % Assi
    \draw[->] (0,0) -- (\xmax,0) node[right] {$e$};
    \draw[->] (0,0) -- (0,\ymax) node[above] {$\bar{p}^*$};
    
    % Etichetta della soglia
    \draw (\RminusV,0) node[below] {$R-V$} -- (\RminusV,0.1);
    \draw (0,1) node[left] {$1$} -- (0.1,1);
    
    % Equilibrio p=0: SEMPRE STABILE (linea blu continua)
    \draw[thick, blue] (0,0) -- (\xmax,0);
    
    % Equilibrio p=1: stabile per e < R-V, instabile per e > R-V
    \draw[thick, blue] (0,1) -- (\RminusV,1);  % stabile
    \draw[dashed, gray] (\RminusV,1) -- (\xmax,1); % instabile

    \def\PminusS{1}
    \def\DeltaVal{4}
    \def\pzero{0.25}
    
    \draw[thick, red, dashed] (0,\pzero) -- (\RminusV,1);
    
    % Legenda
    \draw[blue, thick] (4,0.8) -- (4.3,0.8) node[right] {stable};
    \draw[red, thick, dashed] (4,0.5) -- (4.3,0.5) node[right] {unstable};
    
    % Nota: l'equilibrio interno non esiste per e > R-V
\end{tikzpicture}
\caption{Bifurcation diagram for the coordination game ($\Delta > 0$). For all $e \ge 0$, $\bar{p}=0$ (conflict) is stable. For $e < R-V$, $\bar{p}=1$ (peace) is also stable, and the two stable equilibria are separated by an unstable interior equilibrium (red dashed line). For $e > R-V$, the peaceful equilibrium disappears and only conflict remains.}
\end{figure}

\section{Strategic Command Authority}
\subsection{The time-scale separation model}
In the baseline model the enforcement intensity $e$ was exogenous. We now let a command authority choose $e$ strategically, anticipating the evolutionary response of the frontline. This is a Stackelberg game: the authority is the leader, the population the follower.

 In the Stackelberg model we
implicitly assume that the command authority commits to a fixed
enforcement level $e$ that remains constant while the frontline
population evolves to its long-run equilibrium $\bar{p}^*(e)$. This
slow-evolutionary / fast-command separation is standard in
hierarchical evolutionary models \cite{Sandholm2010} and can be
interpreted as the authority setting institutional rules (e.g., rules of
engagement, disciplinary codes) on a strategic timescale, while tactical
behavior adapts faster through imitation and social learning.

\subsection{Stackelberg Problem with Equilibrium Selection}

We now consider a command authority that chooses enforcement $e$ anticipating the evolutionary response of the frontline. Unlike the previous analysis, the coordination game with $\Delta > 0$ gives rise to two stable pure equilibria ($\bar{p}=0$ and $\bar{p}=1$) for $e < R-V$, and a unique stable equilibrium ($\bar{p}=0$) for $e > R-V$. To obtain a well-defined prediction, we assume that the population is initially at the peaceful equilibrium $\bar{p}=1$,  a natural status quo in contexts where local truces have emerged. Assuming local truces are the
\emph{pre-existing} state and that the commander's decision is whether to preserve or destroy them is key in our model and consistent with historical episodes such as the Christmas Truce, where
truces arose spontaneously before command intervention. An alternative model would assume the
population starting at $\bar{p}=0$, i.e., full conflict, but that case yields trivially that peace cannot emerge without an external shock. The commander, as a Stackelberg leader, can then decide whether to preserve this state or to increase enforcement beyond the threshold $R-V$, thereby destabilizing peace and triggering a transition to conflict.

The commander's payoff is
\[
U_C(e) = \alpha\bigl(1 - \bar{p}^*(e)\bigr) + \beta e - \frac{c}{2}e^2,
\]
where $\bar{p}^*(e)$ now depends on the regime:
\begin{itemize}
    \item If $e \le R-V$, the peaceful equilibrium $\bar{p}=1$ remains stable, so $\bar{p}^*(e) = 1$.
    \item If $e > R-V$, the peaceful equilibrium loses stability and the population converges to $\bar{p}=0$, so $\bar{p}^*(e) = 0$.
\end{itemize}
Thus
\[
U_C(e) = 
\begin{cases}
\beta e - \dfrac{c}{2}e^2, & e \le R-V, \\[8pt]
\alpha + \beta e - \dfrac{c}{2}e^2, & e > R-V.
\end{cases}
\]

Both branches are concave quadratics. Their unconstrained maxima occur at $e = \beta/c$, provided this point lies in the respective interval. Let $e^*_p = \min(\beta/c, R-V)$ denote the best feasible choice in the peace regime, and $e^*_c = \max(\beta/c, R-V)$ the best in the conflict regime. The globally optimal enforcement $e^*$ is the one that yields the higher utility between $U_C(e^*_p)$ and $U_C(e^*_c)$.

\begin{proposition}[Necessary and sufficient
condition for peace]
\label{prop:peace}
The command authority prefers \emph{peace} (i.e.\ $e^* \leq R-V$) if and only if
\[
U_C(e^*_p) \;\geq\; U_C(e^*_c),
\]
that is
\[
  \alpha \;\leq\; \frac{c}{2}\bigl[(e^*_c)^2 - (e^*_p)^2\bigr]
                - \beta\bigl(e^*_c - e^*_p\bigr).
\]
In particular, peace is preferred whenever $\alpha = 0$, and conflict is preferred whenever $\alpha$ exceeds $\frac{c}{2}\bigl[(e^*_c)^2 - (e^*_p)^2\bigr]
                - \beta\bigl(e^*_c - e^*_p\bigr)$. Reducing $\alpha$, reducing $\beta$, or increasing $c$ all shift this inequality in favor of
peace.
\end{proposition}

\subsection{Outcomes}
Four qualitatively different outcomes can emerge, depending on parameters:
\begin{enumerate}
    \item \textbf{Stable peace with low enforcement}: if $\beta/c < R-V$ and $U_C(\beta/c) > U_C(R-V)$, the commander chooses $e^* = \beta/c$ and peace persists.
    \item \textbf{Peace at the threshold}: if $\beta/c > R-V$ but $U_C(R-V) > U_C(\beta/c)$, the commander chooses $e^* = R-V$, remaining in peace but pushing enforcement to its maximum sustainable level.
    \item \textbf{Transition to conflict}: if $\beta/c < R-V$ and $U_C(R-V) > U_C(\beta/c)$, the commander finds it optimal to set $e^* = R-V$ (or slightly above), deliberately destabilizing peace and moving the population to the conflict equilibrium.
    \item \textbf{Full conflict with high enforcement}: if $\beta/c > R-V$ and $U_C(\beta/c) > U_C(R-V)$, the commander chooses $e^* = \beta/c$ and the system operates in the conflict regime.
\end{enumerate}

This formulation captures the strategic tension inherent in hierarchical control: the commander may either ``accept'' peace with moderate enforcement or ``provoke'' conflict by exceeding the threshold $R-V$, trading off the direct benefits of enforcement ($\beta$) against the cost $c$ and the value of conflict ($\alpha$). The model thus predicts when a rational authority will deliberately undermine local truces and this is a key insight for understanding the persistence of war.

\newpage

\section{Policy Implications: How to Induce a Peaceful Choice}

The analysis above shows that a command authority with a direct taste for enforcement ($\beta>0$) may choose either to preserve peace (by setting $e \le R-V$) or to provoke conflict (by setting $e > R-V$), depending on the relative magnitudes of $\alpha$, $\beta$, $c$, and the threshold $R-V$. From a policy perspective, the question is: how can external actors (international organizations, domestic constraints, or institutional reforms) alter the commander's incentives so that peace becomes the optimal choice?

\subsection{Policy Levers}

Three channels directly affect the commander's payoff, based on Proposition 1:
\begin{enumerate}
    \item \textbf{Reducing the benefit of conflict ($\alpha$).} International sanctions, diplomatic pressure, post-conflict accountability mechanisms, or the promise of political integration can lower the commander's payoff from sustained fighting. A lower $\alpha$ reduces the attractiveness of the conflict regime relative to peace.
    \item \textbf{Reducing the direct benefit of enforcement ($\beta$).} This is the most direct channel. If external actors can tax, confiscate, or otherwise appropriate the rents that commanders derive from exercising control, $\beta$ decreases. This could take the form of asset freezes, travel bans, international criminal prosecution, or conditioning foreign aid on reductions in internal repression.
    \item \textbf{Increasing the cost of enforcement ($c$).} Raising the marginal cost of enforcement — for example, through monitoring, reporting requirements, independent oversight, or by making repression more costly diplomatically — makes any positive $e$ more expensive and shifts the optimal enforcement downward.
\end{enumerate}

\subsection{How These Levers Affect the Commander's Choice}

The commander's optimal decision is determined by comparing the maximal utility achievable in the peace regime, $U_p^* = \max_{0 \le e \le R-V} (\beta e - \frac{c}{2}e^2)$, with the maximal utility in the conflict regime, $U_c^* = \max_{e \ge R-V} (\alpha + \beta e - \frac{c}{2}e^2)$. Peace is chosen if and only if $U_p^* \ge U_c^*$.

\begin{figure}[htbp]
\centering
\begin{tikzpicture}[scale=0.65, transform shape, every node/.style={font=\small}]
    % Parametri comuni (valori illustrativi)
    \def\RminusV{3}
    \def\ymax{4}
    \def\xmax{6}
    
    % Panel (a): Baseline conflict regime
    \begin{scope}[xshift=-5cm]
        \draw[->] (0,0) -- (\xmax,0) node[right] {$e$};
        \draw[->] (0,0) -- (0,\ymax) node[above] {$U_C$};
        \draw[dashed] (\RminusV,0) node[below] {$R-V$} -- (\RminusV,\ymax-0.5);
        
        % Peace branch (e <= R-V)
        \draw[thick,blue] plot[domain=0:\RminusV,samples=20] (\x,{1.5*\x - 0.3*\x*\x});
        % Conflict branch (e >= R-V)
        \draw[thick,red] plot[domain=\RminusV:5.5,samples=20] (\x,{2.5 + 1.5*\x - 0.3*\x*\x});
        
        \fill[red] (4.5,3.8) circle (2pt) node[above] {$e^*$};
        \node at (3,\ymax-0.5) {(a) Conflict prevails};
    \end{scope}
    
    % Panel (b): Moderate incentives
    \begin{scope}[xshift=0cm]
        \draw[->] (0,0) -- (\xmax,0) node[right] {$e$};
        \draw[->] (0,0) -- (0,\ymax) node[above] {$U_C$};
        \draw[dashed] (\RminusV,0) node[below] {$R-V$} -- (\RminusV,\ymax-0.5);
        
        \draw[thick,blue] plot[domain=0:\RminusV,samples=20] (\x,{1.2*\x - 0.3*\x*\x});
        \draw[thick,red] plot[domain=\RminusV:5.5,samples=20] (\x,{1.8 + 1.2*\x - 0.3*\x*\x});
        
        \fill[blue] (2,1.2) circle (2pt) node[above] {$e^*$};
        \node at (3,\ymax-0.5) {(b) Peace becomes competitive};
    \end{scope}
    
    % Panel (c): Peace induced
    \begin{scope}[xshift=5cm]
        \draw[->] (0,0) -- (\xmax,0) node[right] {$e$};
        \draw[->] (0,0) -- (0,\ymax) node[above] {$U_C$};
        \draw[dashed] (\RminusV,0) node[below] {$R-V$} -- (\RminusV,\ymax-0.5);
        
        \draw[thick,blue] plot[domain=0:\RminusV,samples=20] (\x,{1.0*\x - 0.3*\x*\x});
        \draw[thick,red] plot[domain=\RminusV:5.0,samples=20] (\x,{1.0 + 1.0*\x - 0.3*\x*\x});
        
        \fill[blue] (1.67,0.83) circle (2pt) node[above] {$e^*$};
        \node at (3,\ymax-0.5) {(c) Peace prevails};
    \end{scope}
\end{tikzpicture}
\caption{Effect of policy interventions on the commander's optimal enforcement. (a) Baseline with high $\alpha,\beta$ leads to conflict ($e^* > R-V$). (b) Moderate incentives make peace competitive. (c) Strong incentives induce peace ($e^* \le R-V$).}
\label{fig:policy_shifts}
\end{figure}
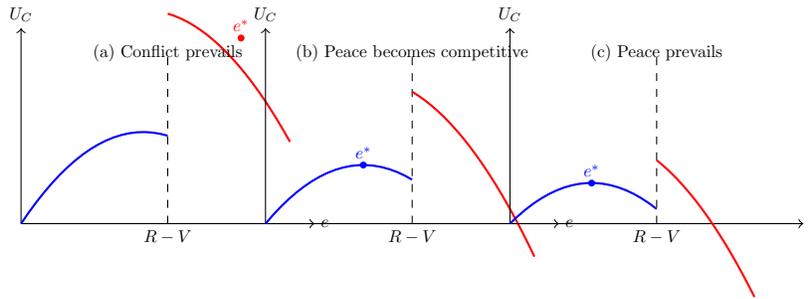

Figure~\ref{fig:policy_shifts} illustrates how policy interventions shift these maxima. In panel (a), the baseline parameters favor conflict: $U_c^* > U_p^*$ and the commander chooses $e^* > R-V$. Panel (b) shows the effect of moderate incentives: reducing $\alpha$ or $\beta$, or increasing $c$, brings the two maxima closer, possibly making peace competitive. Panel (c) depicts the ultimate goal: $U_p^* > U_c^*$, so the commander optimally chooses an enforcement level $e^* \le R-V$, sustaining peace.

\subsection{Policy Instruments in Practice}

Several concrete instruments can implement these shifts:
\begin{itemize}
    \item \textbf{International sanctions and prosecutions.} By targeting the personal wealth and freedom of commanders (e.g., through the International Criminal Court, asset freezes, travel bans), external actors directly reduce $\beta$.
    \item \textbf{Conditional aid and peace conditionalities.} Making financial or military assistance conditional on observable reductions in enforcement (lower $e$) effectively raises $c$ by attaching a cost to enforcement.
    \item \textbf{Post-conflict power-sharing and integration.} Guaranteeing commanders a political future, amnesty, or economic opportunities after peace reduces their stake in continued conflict, lowering $\alpha$.
    \item \textbf{Transparency and monitoring.} Independent monitors can report on troop treatment and enforcement activities, increasing the reputational cost of repression (higher $c$).
    \item \textbf{Strengthening the peaceful equilibrium.} Although not directly in the commander's payoff, policies that increase $R$ (the reward for mutual cooperation) or decrease $V$ (the temptation to attack) raise the threshold $R-V$, expanding the range of enforcement levels compatible with peace. This can be achieved through confidence-building measures, economic integration, or social programs that make peace more valuable for soldiers.
\end{itemize}

Cases 1 and 4 are the interior solutions where the unconstrained optimum $\beta/c$ lies within $R-V$; cases 2 and 3 are corner solutions where the binding constraint is the threshold itself. Case 3 is the most policy-relevant, as it describes a commander who, in the absence any conflict motivation ($\alpha = 0$), would prefer low enforcement ($\beta/c < R-V$), but the prospect of capturing the full benefit of conflict, which only becomes available once peace is destroyed, can make crossing that line worth.

The model thus provides a clear rationale for a multi-pronged strategy: sustainable peace requires not only changing soldiers' incentives, but also restructuring the incentives of those in power. By making enforcement less personally beneficial, more costly, and the benefits of conflict smaller, external actors can tilt the commander's calculus toward peace.

Notice that a positive enforcement level in the peace regime is not paradoxical once we recognize that $\beta>0$ represents the commander's private benefit from the act of enforcement itself, such as prestige, control over resources, or institutional power, which can make $e>0$ desirable even when it does not alter frontline behavior. This mechanism finds empirical support across diverse historical contexts: in post-Soviet Afghanistan, local warlords maintained checkpoints and armed followings primarily to extract rents and demonstrate authority rather than to fight active enemies (\cite{Giustozzi}). Similarly, Somali faction leaders during the 1990s continued to tax populations and control ports in relatively peaceful areas, using enforcement as a source of revenue and patronage (\cite{Marten}). In Colombia, paramilitary commanders preserved territorial control to secure drug routes and political influence long after active combat had ceased in their regions (\cite{Duncan}). Historical parallels abound: late Roman generals and landowners maintained private armed retainers (\textit{bucellarii}) as symbols of status and instruments of local power even in provinces far from active fronts (\cite{halsall}), while medieval European lords exercised judicial and military authority over their lands collecting fines, demanding labor, and holding courts as a direct source of income and prestige, regardless of external wars (\cite{Bloch}). These examples illustrate that enforcement often serves private purposes beyond its ostensible function, validating the inclusion of $\beta>0$ in the commander's objective function.

\subsection{The Role of Civil Society}

The model also illuminates the potential of civil society to resolve the classic puzzle of how a minority in power can impose war on an unwilling majority (a theme captured in Trilussa's 1914 poem "Ninna nanna della guerra," which contrasts the interests of those who profit from conflict with the suffering of those who fight). While peace ($\bar{p}=1$) is a stable equilibrium, it may be undermined by a commander with private incentives ($\beta>0$). Civil society can counteract this through several channels.
First, through \emph{Monitoring and transparency:}: by documenting and publicizing instances of enforcement against peaceful soldiers, civil society organizations raise the reputational cost of repression, effectively increasing the parameter $c$ in the commander's objective function.
Second, \emph{Social pressure and normative change} may play a very imoortant role. In fact, when peace becomes a widely shared norm, it raises the psychological and social cost of defection for the commander, again increasing $c$ or reducing the private benefit $\beta$ of exercising power.
Third, Civil society can create \emph{alternative institutions}, that is parallel structures of dispute resolution and community governance that make the peaceful equilibrium more robust, effectively increasing $R$ (the reward for mutual cooperation) or reducing $V$ (the temptation to attack), thereby expanding the range $R-V$ within which peace remains stable.
Finally, civil society may push \emph{Collective action and non-violent resistance} . When soldiers and civilians coordinate to resist enforcement, they can raise the cost of repression to prohibitive levels, making $e>0$ too expensive for the commander.

In this framework, peace becomes robust precisely when bottom-up social mobilization raises the costs of enforcement to the point where the commander's private calculus favors the peaceful equilibrium. The model thus provides a formal foundation for the intuition that sustainable peace requires not only top-down institutional design but also an active civil society that makes war too costly for those who would impose it — transforming people in an organized force capable of resisting top-down war.

 \section{Discussion and Policy Implications}

The analysis developed in this paper has direct implications for the institutional design of conflict management and peace enforcement mechanisms. A central insight of the model is that local cooperation among adversaries can emerge endogenously through repeated interaction and evolutionary adaptation, nevertheless it still remains dynamically fragile when higher-level authorities retain incentives to sustain hostilities. 

Historically, this mechanism is vividly illustrated by the Christmas Truce of 1914. In that episode, soldiers on opposing sides of the Western Front spontaneously converged toward mutual restraint, exchanging goods, coordinating ceasefires, and jointly retrieving casualties. From the perspective of the model, this corresponds to a temporary shift in the population state toward a high-cooperation equilibrium driven by repeated local interaction and reciprocal expectations. However, the intervention of military command rapidly eliminated these cooperative outcomes by imposing strict disciplinary measures, reassigning units, and redefining engagement rules. In terms of the model, these actions increased the effective enforcement parameter, shifting the system out of the basin of attraction of cooperative equilibria and restoring full conflict as the only stable outcome. The episode thus exemplifies how hierarchical incentives can override decentralized cooperation even when the latter is individually rational and evolutionarily stable in the absence of centralized control.

The same structural logic applies to contemporary conflicts. In several modern war zones, ethnographic and journalistic accounts document localized ceasefires, tacit coordination over humanitarian corridors, and informal rules of engagement aimed at minimizing casualties. Nevertheless, these cooperative practices rarely evolve into lasting peace. The model provides a parsimonious explanation for this empirical regularity by identifying a conflict of interest between frontline agents and command authorities. While combatants directly internalize the material and psychological costs of fighting, political and military leadership may face incentives that favor prolonged conflict, including strategic bargaining advantages, domestic political consolidation, or control over economic rents associated with warfare. When these incentives dominate, command authorities optimally maintain enforcement levels that suppress cooperative dynamics at the local level.

This theoretical result has important implications for peace-building strategies. 

From an institutional perspective, the model underscores the importance of external enforcement and accountability mechanisms. International monitoring missions, judicial prosecution of ceasefire violations, and conditionality frameworks linking economic or political benefits to verified reductions in hostilities can be interpreted as devices that increase the marginal cost of enforcement for command authorities. In the formal structure of the model, such mechanisms effectively reduce the feasible range of the enforcement parameter, thereby expanding the region of parameter space in which peaceful equilibria are stable. 

The model also provides insight into the recurrent fragility of peace processes in post-conflict societies. Even when frontline violence subsides and local cooperation becomes widespread, the persistence of political structures that reward coercion or militarization creates a latent instability. Under such conditions, peace remains a metastable state vulnerable to shocks, leadership changes, or strategic provocations. Durable peace therefore requires not only demobilization and reconciliation at the micro level, but also constitutional or international constraints that limit the ability of elites to extract rents from renewed conflict.

The model offers also a lens through which to interpret persistent conflict dynamics observed in contemporary geopolitics. When enforcement is cheap and the private benefit of conflict is large ($\alpha$ high, $c$ low), our analysis predicts an endless war equilibrium in which a rational authority actively maintains hostilities even when frontline soldiers would prefer truce. This prediction is consistent with a broad class of historical and contemporary cases in which leadership incentives diverge from frontline preferences. On the contrary, when enforcement becomes expensive (higher $c$) due to economic pressures or electoral concerns, even leaders who derive intrinsic utility from wielding power may find continuation less attractive. Similarly, protracted conflicts such as the war in Ukraine exemplify the "endless war" equilibrium that emerges when the central authority's incentives align with prolonged hostilities.  The model thus implies that sustainable peace requires restructuring the incentives of power-holders themselves — raising their costs ($c$), reducing their private benefits from enforcement ($\beta$), and making conflict strategically untenable for decision-makers, not only for frontline soldiers.

 Effective peace policy, according to the model, must therefore be evaluated not only by its capacity to foster cooperation at the grassroots level, but by its ability to impose credible limits on the incentives of command authorities to disrupt that cooperation. Only when such limits are in place does the cooperative equilibrium become robust to strategic manipulation and capable of persisting over time.
Our analysis also suggests that peace, as a stable equilibrium, can prevail when bottom-up social mobilization raises the costs of enforcement sufficiently to alter the commander's private calculus. Civil society organizations, through monitoring, norm-building, and collective action, effectively increase the parameter the cost of enforcement) or reduce the private benefit from exercising power.

\section{Conclusions}

This paper has developed an evolutionary game-theoretic model of cooperation in conflict environments that explicitly incorporates hierarchical enforcement and command-level incentives. By embedding decentralized interaction within a structure of top-down control, the analysis explains how local cooperation may emerge endogenously but still fail to persist when higher authorities retain incentives to sustain hostilities.

The results identify enforcement intensity as the key parameter governing long-run outcomes. When command-level incentives favor conflict, cooperative equilibria are eliminated and full conflict becomes globally stable. By contrast, when institutional or legal constraints limit the capacity to penalize local cooperation, peaceful equilibria become stable and self-reinforcing. This mechanism reconciles the empirical coexistence of spontaneous truces and prolonged wars by showing that cooperation can be evolutionarily viable but institutionally fragile.

From a policy perspective, the analysis implies that sustainable peace requires not only bottom-up cooperation but also binding constraints on the strategic incentives of command authorities. 
Overall, the paper highlights the central role of institutional design in transforming transient cooperation into durable peace and offers a theoretical foundation for policies aimed at constraining command-level incentives rather than solely fostering individual-level reconciliation.
The model further includes the possibility that a sustainable peace can be supported  not only through institutional constraints on the commander, but also thanks to an organized civil society capable of raising the costs of enforcement until war becomes too expensive even for those in power.

The model presented in this paper offers a tractable framework for analyzing the interplay between decentralized evolutionary dynamics and hierarchical incentives in conflict settings. Several extensions could enrich the analysis and connect it more closely with empirical phenomena.

First, incorporating stochastic effects and finite population sizes would allow us to study how demographic fluctuations can lead to fixation or extinction of cooperative strategies, particularly near the boundaries where deterministic forces are weak. This is especially relevant for small frontline units where chance events may play a significant role.

Second, empirical calibration of the model using historical or experimental data could help identify the parameter regions most relevant to actual conflicts, and potentially inform policy interventions aimed at fostering local truces.

These extensions notwithstanding, the core message of the paper remains: hierarchical incentives interact with local strategic adaptation in subtle ways, and a rational authority may paradoxically avoid using its enforcement power when it would increase frontline cooperation.

\end{document}